\documentclass{article}
\usepackage{amsmath}
\usepackage{amssymb}
\usepackage{latexsym}
\newtheorem{lemma}{Lemma}

\newtheorem{Proposition}{Proposition}
\newtheorem{Def}{Definition}

\def\bbbr{{\rm I\!R}} 

\def\cA{{\cal A}}

\def\cC{{\cal C}}
\def\cH{{\cal H}}

\def\cN{{\cal N}}

\def\PN{{\cal P} ({\cal N} )}
\def\c{{\bot}} 
\def\vagy{\vee}
\def\es{\wedge}
\def\C{$C^{\ast}$--}

\begin{document}

\title{ Remarks on Causality in \\
Relativistic Quantum Field Theory }

\author{{Mikl\'os R\'edei} \\
Department of History and Philosophy of Science  \\
Lor\'and E\"otv\"os University\\
P.O. Box 32, H-1518 Budapest 112,
Hungary\thanks{e-mail: redei@ludens.elte.hu}\\
\vphantom{X}\\
{Stephen J.\ Summers } \\
Department of Mathematics, University of Florida,\\
Gainesville FL 32611, USA\thanks{e-mail: sjs@math.ufl.edu}}

\date{}
\maketitle
\begin{center}
{\bf Abstract}\\[12pt]
\end{center}

{It is shown that the correlations predicted by relativistic quantum field
theory in locally normal states between
projections in local von Neumann algebras $\cA(V_1),\cA(V_2)$ associated
with spacelike separated spacetime regions $V_1,V_2$ have a (Reichenbachian)
common cause located in the union of the backward light cones of $V_1$ and
$V_2$. Further comments on causality and independence in quantum field
theory are made.}

\section{Introduction}

      Algebraic quantum field theory (AQFT) (cf.
\cite{Haag1992}) predicts
correlations between projections $A,B$ lying in von Neumann algebras
$\cA(V_1),\cA(V_2)$ associated with spacelike separated spacetime
regions $V_1,V_2$. According to {\em Reichenbach's Common Cause
Principle} (cf. \cite{Salmon1984}) if two events $A$ and $B$ are
correlated, then the correlation between $A$ and $B$ is either due to
a direct causal influence connecting $A$ and $B$, or there is a third
event $C$ which is a common cause of the correlation. The latter means
that $C$ satisfies four simple probabilistic conditions which together
imply the correlation in question.

     The correlations predicted by AQFT lead naturally to the question
of the status of Reichenbach's Common Cause Principle within AQFT.  If
the correlated projections belong to algebras associated with
spacelike separated regions, a direct causal influence between them is
excluded by the theory of relativity. Consequently, compliance of AQFT
with Reichenbach's Common Cause Principle would mean that for every
correlation between projections $A$ and $B$ lying in von Neumann
algebras associated with spacelike separated spacetime regions
$V_1,V_2$, there must exist a projection $C$ possessing the
probabilistic properties which qualify it to be a Reichenbachian
common cause of the correlation between $A$ and $B$. However, since
observables and hence also the projections in AQFT must be localized,
one also has to specify the spacetime region $V$ with which the von
Neumann algebra $\cA(V)$ containing the common cause $C$ is
associated. Intuitively, the region $V$ should be disjoint from both
$V_1$ and $V_2$ but should not be causally disjoint from them, in
order to leave room for a causal effect of $C$ on the correlated
events. There are three natural candidates for such a region $V$: the
intersection of the backward light cones of $V_1$ and $V_2$
($cpast(V_1,V_2)$, see (\ref{C})), the intersection of the backward
light cones of every point in $V_1$ and $V_2$
($spast(V_1,V_2)$, see (\ref{S})) and the union of the backward light
cones of $V_1$ and $V_2$ ($wpast(V_1,V_2)$, see (\ref{W})).
The requirement that the common cause belongs to local algebras
associated with spacetime regions $spast(V_1,V_2)$, $cpast(V_1,V_2)$
and $wpast(V_1,V_2)$, leads to three different specifications of
Reichenbach's Common Cause Principle in AQFT, called Strong Common
Cause Principle, Common Cause Principle and Weak Common Cause
Principle, respectively (Definition \ref{defccqft}). Since
$spast(V_1,V_2)=\emptyset$ if $V_1$ and $V_2$ are complementary wedge
regions and AQFT predicts correlations between projections localized
in complementary wedges (see below), the Strong Common Cause Principle
fails in AQFT. Whether the Common Cause Principle holds is still an
open problem.

     We show that the Weak Common Cause Principle holds for every
local system $(\cA(V_1),\cA(V_2),\phi)$ with a locally normal and
locally faithful state $\phi$ and suitable, bounded spacelike
separated spacetime regions $V_1,V_2$, if a net $\{\cA(V)\}$
satisfies some standard, physically natural
assumptions as well as the so--called {\em local primitive causality}
condition (Definition \ref{primitive}).  Such states include the
states of physical interest in vacuum representations for relativistic
quantum field theories on Minkowski space. We shall interpret our main
result, Proposition \ref{main}, as a clear demonstration that AQFT is
a causally rich enough theory to comply with the Weak Common Cause
Principle -- and possibly also with the Common Cause Principle without
the qualification ``weak".

     In the next section we shall specify the assumptions
and some immediate consequences of these assumptions
needed in the proof of the main result. In Section 3 the definitions
of the Reichenbach's Common Cause Principles for AQFT are given,
followed by the main result.  In the last section we shall make some
further comments about our results.

\section{Spacelike Correlations in Quantum Field Theory}

     Throughout the paper $\{ \cA(V)\}$ denotes a net of local von
Neumann algebras (indexed by the open, bounded subsets $V$ of
Minkowski space $M$) satisfying the standard axioms of (i) isotony,
(ii) Einstein causality, (iii) relativistic covariance and acting on a
Hilbert space $\cH$ carrying an irreducible vacuum representation of
the net.  The representation of the Poincare group is therefore (iv)
implemented by a (strongly continuous) unitary representation $U$
satisfying the spectrum condition and having a distinguished invariant
vector $\Omega \in \cH$ representing the vacuum state. In addition to
(i)-(iv), we also assume (v) weak additivity: for any nonempty open
region $V$, the set of operators $\cup_{x \in {\bbbr}^4} \cA(V+x)$ is
dense in $\cup_{V \subset M} \cA(V)$ (in the weak operator topology).
(For further discussion of these axioms, see \cite{Haag1992} and
\cite{Horuzhy1990}.)

     An immediate consequence of assumptions (i)--(v) is that we may
employ the following result of Borchers:

\begin{Proposition} {\rm \cite{Borchers}}
Under the assumptions (i)--(v), for any
nonempty open region $V$, the set of vectors $\cA(V)\Phi$ is dense
in $\cH$, for any vector $\Phi$ which is analytic for the energy.
\end{Proposition}

     Note that any vector $\Phi$ with finite energy content, in
particular the vacuum, is analytic for the energy.  And since no
preparation of a quantum system which can be carried out by man can
require infinite energy, it is evident that (convex combinations of)
states induced by such analytic vectors include all of the physically
interesting states in this representation.

     Note further that assumption (ii) entails that such
vectors are also separating ({\it i.e.} $X \in \cA(V)$ and $X\Phi
= 0$ imply $X = 0$) for all algebras $\cA(V)$ such that $V'$ is
nonempty. (Here $V'$ denotes the causal complement and $V''=(V')'$
denotes the causal completion of a convex spacetime region $V$.)
Hence, for each bounded region $V$ (convex combinations of) the
states $\phi$ induced by analytic vectors are faithful on each
such algebra $\cA(V)$ ({\it i.e.} $X \in \cA(V)$ and $\phi(XX^*) =
0$ imply $X = 0$). Such states are said to be locally faithful. We
emphasize: given assumptions (i)--(v), all physically interesting
states in the vacuum representation will be locally faithful.

     We shall also assume that (vi) the net $\{\cA(V)\}$ and state
$\phi$ have a nontrivial scaling limit, either in the sense of
Fredenhagen \cite{Fredenhagen} or in the sense of Buchholz and Verch
\cite{Buchholz-Verch}.
This assumption has been verified in many concrete models and is
expected to hold in any renormalizable quantum field theory with an
ultraviolet fixed point, hence in all asymptotically free theories.
The role of this physically motivated assumption in our argument is
to provide information about the type of the local algebras $\cA(V)$
which can occur.

\begin{Def}\label{primitive}
The net $\{\cA(V)\}$ is said to satisfy the {\em local primitive causality}
condition if $\cA(V'')=\cA(V)$ for every nonempty convex region $V$.
\end{Def}

\noindent  Local primitive causality postulates that the quantum field
undergoes a
hyperbolic propagation within lightlike characteristics \cite{Haag-Schroer}.
(See the discussion in Section 4 for further insight into the nature of this postulate.)
Our final assumption is that the net satisfies the local primitive causality.
This assumption does not follow from
assumptions (i)--(vi) (see \cite{Garber}). However, this condition
has been verified in many concrete models.

For spacetime point $x\in M$ let $V_+(x)$ ($V_-(x)$) denote the open forward
(backward) light cones with apex $x$.
If $x \in V_+(y)$ then
$V_-(x) \cap V_+(y)$ is called a double cone.  If $V$ is a double cone, then $V$ and $V'$ are nonempty and
$V = V''$. The wedge regions are Poincar\'e
transforms of the basic wedge
$$W_R = \{ (x_0,x_1,x_2,x_3) \in M \mid x_1 > \vert x_0 \vert \} \quad .$$
Note that wedges are unbounded sets and $W = W''$ for every wedge $W$.
Moreover, if $W$ is a wedge, then so is $W'$. As shown in \cite{Fredenhagen},
assumptions (i)--(vi) entail that the algebra $\cA(V)$ is type $III$
whenever $V$ is a double cone or a wedge.

     We shall need some definitions and results concerning the
independence of local algebras.\footnote{For the origin and a detailed
analysis of the interrelation of these and other notions of
statistical independence, see the review \cite{Summers1990} and
Chapter 11 in \cite{konyvem} --- for more recent results, see
\cite{Florig-Summers1997}\cite{Hamhalter}.}  A pair $(\cA_1,\cA_2)$ of
\C subalgebras of the \C algebra $\cC$ has the Schlieder property if
$XY\not=0$ for any $0\not= X\in\cA_1$ and $0\not=Y\in\cA_2$.
Given assumptions (i)--(v), $(\cA(V_1),\cA(V_2))$ has the
Schlieder property for all spacelike separated double cones
or wedges \cite{Summers1990}.

     A pair $(\cA_1,\cA_2)$ of such algebras is called \C independent
if for any state $\phi_1$ on $\cA_1$ and for any state $\phi_2$ on
$\cA_2$ there exists a state $\phi$ on $\cC$ which extends both
$\phi_1$ and $\phi_2$.  Under assumptions (i)--(v), algebras
associated with spacelike separated double cones are \C independent,
since they form a mutually commuting pair of
algebras satisfying the Schlieder property, which in this context is
equivalent with \C independence \cite{Roos}.

     Two von Neumann subalgebras $\cN_1,\cN_2$ of the von Neumann
algebra $\cN$ are called logically independent
\cite{Redei1995a}\cite{Redei1995d} if $A\es B\not=0$ for
any  projections $0\not=A\in\cN_1$, $0\not=B\in\cN_2$. If
$\cN_1,\cN_2$ is a mutually commuting pair,
then \C independence and logical independence are equivalent
\cite{konyvem}.\footnote{If $\cN_1,\cN_2$ do not mutually
commute, then \C independence is strictly weaker than logical
independence \cite{Hamhalter}.} So we
conclude:

\begin{lemma} \label{logind} Assumptions (i)--(v) entail that the pair
$(\cA(V_1),\cA(V_2))$ is logically independent for any spacelike
separated double cones or wedges $V_1,V_2$.
\end{lemma}

     Let $V_1$ and $V_2$ be two spacelike separated spacetime
regions and $A\in\cA(V_1)$ and $B\in\cA(V_2)$ be two projections. If $\phi$ is
a state on $\cA(V_1\cup V_2)$ and
\begin{equation}\label{supcorr}
\phi(A\es B)>\phi(A)\phi(B) \quad ,
\end{equation}
then we say that there is {\em superluminal
(or spacelike) correlation} between $A$ and $B$ in the state $\phi$.
We now explain why such correlations are common when assumptions
(i)--(v) hold.

     The ubiquitous presence of superluminal correlations is one of the
consequences of the generic violation of Bell's inequalities in AQFT.
To make this clear, recall (cf. \cite{Summers-Werner1985}) that  the Bell correlation
$\beta(\phi,\cN_1,\cN_2)$ between two commuting von Neumann subalgebras $\cN_1,\cN_2$ of the von
Neumann algebra $\cN$
in state $\phi$ on $\cN$ is defined by
\begin{equation} \label{defbellcorr}
\beta(\phi,\cN_1,\cN_2)\equiv
\sup \;
\frac{1}{2}\phi(X_1(Y_1+Y_2)+X_2(Y_1-Y_2)) \quad ,
\end{equation}
where the supremum in (\ref{defbellcorr}) is taken over all self--adjoint
contractions $X_i\in\cN_1,Y_j\in\cN_2$.  It can
be shown \cite{Cirelson}\cite{Summers-Werner1987b} that
$\beta(\phi,\cN_1,\cN_2)\leq\sqrt{2}$. The Clauser--Holt--Shimony--Horne
version of Bell's inequality in this notation reads:
\begin{equation}\label{defbellineq}
 \beta(\phi,\cN_1,\cN_2)\leq 1
\quad ,
\end{equation}
and a state $\phi$ for which $\beta(\phi,\cN_1,\cN_2)> 1$ is called {\em Bell
correlated}. It is known \cite{Summers-Werner1987b} that if $\phi$ is a product state across the algebras
$\cN_1,\cN_2$ ({\it i.e.}, if $\phi(XY) = \phi(X)\phi(Y)$, for all
$X \in \cN_1$ and $Y \in \cN_2$), then $\beta(\phi,\cN_1,\cN_2)=1$.
This in turn implies the next lemma.

\begin{lemma}\label{nonproduct} {\rm \cite{RS}}
Let $\cN_1$ and $\cN_2$ be commuting subalgebras of the von
Neumann algebra $\cN$ and let
$\phi$ be a normal state on $\cN$ which is not a product state across the
algebras $\cN_1,\cN_2$. Then there exist projections $A \in \cN_1$
and $B \in \cN_2$ such that $\phi(A\es B) > \phi(A)\phi(B)$.
\end{lemma}

     There are many situations in which
$\beta(\phi,\cA(V_1),\cA(V_2)) = \sqrt{2}$
(cf. \cite{Summers-Werner1987b}\cite{Summers-Werner1987a}\cite{Summers-Werner1988}).
We recall a recent result by Halvorson and Clifton. Let the symbol
$\cN_1\vee\cN_2$ denote the smallest von Neumann algebra containing both
$\cN_1$ and $\cN_2$.

\begin{Proposition}\label{manycorr}  {\rm \cite{Halvorson-Clifton}}
If $(\cN_1,\cN_2)$ is a pair of commuting type $III$ von Neumann
algebras acting on the Hilbert space $\cH$ and having the Schlieder property,
then the set of unit vectors which induce Bell correlated states on
$\cN_1,\cN_2$ is open and dense in the unit sphere of $\cH$. Indeed,
the set of normal states on $\cN_1\vee\cN_2$ which are Bell correlated
on $(\cN_1,\cN_2)$ is norm dense in the normal state space of $\cN_1\vee\cN_2$.
\end{Proposition}
We see then that, given the assumptions (i)--(vi),
for any spacelike separated double cones or wedges $V_1,V_2$, the pair
$(\cA(V_1),\cA(V_2))$ satisfies the hypothesis of Prop. \ref{manycorr}.
So, ``most'' normal states on such pairs of algebras manifest superluminal
correlations (\ref{supcorr}). Hence, superluminal correlations abound in AQFT,
and the question posed in the introduction is not vacuous.

\section{The Notion of Reichenbachian Common Cause in AQFT}

    The following definition is a natural formulation in a noncommutative
probability space $(\PN,\phi)$\footnote{$\PN$ is the set of all projections
in the von Neumann algebra $\cN$.} of the classical notion of common cause
given by Reichenbach (\cite{Reichenbach1956} Section 19).

\begin{Def}\label{defqcc}
Let $A,B\in\PN$ be two commuting projections which are correlated in $\phi$:
\begin{equation}\label{qcorr}
\phi(A\es B)>\phi(A)\phi(B) \quad .
\end{equation}
$C\in\PN$ is a {\em common cause} of the correlation (\ref{qcorr}) if
$C$ commutes with both $A$ and $B$ and the following conditions hold:
\begin{eqnarray}
\phi(A\es B|C)
&=&\phi(A|C)\phi(B|C) \quad ,
\label{AQFTscreen1}\\
\phi(A\es B|C^\c)&=&\phi(A|C^\c)\phi(B|C^\c) \quad ,
\label{AQFTscreen2}\\
\phi(A|C)&>&\phi(A|C^\c) \quad ,
\label{AQFTnagy1}\\
\phi(B|C)&>&\phi(B|C^\c) \quad .
\label{AQFTnagy2}
\end{eqnarray}
($\phi(X|Y)$ denotes the conditional probability $\phi(X|Y)=\phi(X\es Y)/\phi(Y)$.)
\end{Def}
For spacelike separated spacetime
regions $V_1$ and $V_2$ let us define the following regions
\begin{eqnarray}
wpast(V_1,V_2)&\equiv& (BLC(V_1)\setminus V_1)\cup (BLC(V_2)\setminus V_2)\label{W} \quad , \\
cpast(V_1,V_2)&\equiv& (BLC(V_1)\setminus V_1)\cap (BLC(V_2)\setminus V_2)\label{C} \quad , \\
spast(V_1,V_2)&\equiv& \cap_{x\in V_1\cup V_2}BLC(x)\label{S} \quad ,
\end{eqnarray}
where $BLC(V)$ denotes the union of the backward lightcones of every
point in $V$. Region $spast(V_1,V_2)$ consists of spacetime points
{\em each} of which can causally influence {\em every} point in both $V_1$ and
$V_2$; region $cpast(V_1,V_2)$ consists of spacetime points {\em each}
of which can causally influence at least some point in {\em both}
$V_1$ {\em and} $V_2$, and region $wpast(V_1,V_2)$ consists of
spacetime points {\em each} of which can causally influence at least
{\em some} point in {\em either} $V_1$ {\em or} $V_2$.

\begin{Def} \label{defccqft}
Let $\{{\cal A}(V)\}$ be a net of local von Neumann algebras over Minkowski
space. Let $V_1$ and $V_2$ be two spacelike separated spacetime regions, and
let $\phi$ be a locally normal state on the net. If for any pair of
projections $A\in{\cal A}(V_1)$ and $B\in{\cal A}(V_2)$ the inequality
\begin{equation}\label{qftcorr}
\phi(A\wedge B)>\phi(A)\phi(B)
\end{equation}
entails the existence of a projection $C$ in the von Neumann algebra
${\cal A}(V)$ which is a common cause of the correlation (\ref{qftcorr})
in the sense of Definition \ref{defqcc}, then the local system
$(\cA(V_1),\cA(V_2),\phi)$ is said to satisfy the
\begin{eqnarray}
\mbox{{\bf Weak Common Cause Principle } if \ \ \ }
V&\subseteq& wpast(V_1,V_2) \quad , \label{WCCP}\\
\mbox{{\bf Common Cause Principle } if  \ \ \ }
V&\subseteq&  cpast(V_1,V_2) \quad ,   \label{CCP}\\
\mbox{{\bf Strong Common Cause Principle } if\ \ \ }
V&\subseteq& spast(V_1,V_2) \quad . \label{SCCP}
\end{eqnarray}
We say that Reichenbach's Common Cause Principle holds for the net
(respectively holds in the weak or strong sense) iff for every pair of
spacelike separated convex spacetime regions $V_1,V_2$ and every normal
state $\phi$, the Common Cause Principle holds for the local system
$({\cal A}(V_1),{\cal A}(V_2),\phi)$ (respectively in the weak or strong
sense).
\end{Def}
If $V_1$ and $V_2$ are complementary wedges then
$spast(V_1, V_2)=\emptyset$. Since the local von Neumann algebras pertaining
to complementary wedges are known to contain correlated projections (see
\cite{Summers-Werner1988} and \cite{Summers1990}), the {\em Strong}
Reichenbach's Common Cause Principle trivially fails in AQFT.

     The problem of whether the Common Cause Principle holds in AQFT was
raised in \cite{Redei1997}, and the problem is still open. For the
Weak Common Cause Principle we have the following result.

\begin{Proposition}\label{main}
If the net $\{\cA(V)\}$ satisfies conditions (i)--(vi) and local
primitive causality, then every local system $(\cA(V_1),\cA(V_2),\phi)$
with $V_1,V_2$ nonempty convex open sets such that $V_1{}''$ and $V_2{}''$
are spacelike separated double cones and with a locally normal and
locally faithful state $\phi$ satisfies the Weak Common Cause Principle.
\end{Proposition}

\noindent
The proof of Proposition \ref{main} is based on the following two lemmas:

\begin{lemma} \label{suffcond}
Let $\phi$ be a faithful state on a von Neumann algebra $\cN$
containing two mutually commuting subalgebras $\cN_1,\cN_2$ which are
logically independent. Let $A \in \cN_1$ and $B \in \cN_2$ be projections
satisfying (\ref{qcorr}). Then a sufficient condition for $C$ to satisfy
(\ref{AQFTscreen1})-(\ref{AQFTnagy2}) is that the following two conditions
hold:
\begin{eqnarray}
C & < & A\es B \quad , \label{eleg1}  \\
\phi(C)&=&\frac{\phi(A\es B)-\phi(A)\phi(B)}{1-\phi(A\vagy B)} \quad .
\label{eleg2}
\end{eqnarray}
\end{lemma}

\begin{lemma} \label{proj}
Let $\cN$ be a type $III$ von Neumann algebra on a separable
Hilbert space $\cH$, and let $\phi$ be
a faithful normal state on $\cN$. Then for every projection $A\in\PN$ and
every positive real number $0<r<\phi(A)$ there exists a projection $P\in\PN$
such that $P<A$ and $\phi(P)=r$.
\end{lemma}

\noindent Due to space limitations, we must refer the reader to
\cite{RS} for the proof of these assertions.

\section{Final Remarks}

     The local primitive causality condition plays an essential role
in our proof of Prop. \ref{main} but is barely discussed in the
literature.  If $V$ is a convex region, then for any point $x \in V''$,
every inextendible causal curve through $x$ must intersect
$V$. Hence, the values of a classical quantum field satisfying a
hyperbolic equation of motion whose speed of propagation is bounded by
that of light would at every point in $V''$ be completely determined
by its values in $V$.  This well-known state of affairs finds an
analogous expression in quantum field theory in the condition
$\cA(V) = \cA(V'')$. For free quantum fields there is an explicit link
between the mentioned fact about classical fields and the condition
$\cA(V) = \cA(V'')$ --- cf. \cite{Glimm-Jaffe}. For interacting quantum fields,
the link is significantly more indirect, but has been verified in many
concrete models --- see again \cite{Glimm-Jaffe} for references.  For
this reason, workers in AQFT take the condition of local primitive
causality in general as an expression of hyperbolic propagation within
lightlike characteristics.

     The validity of the local primitive causality condition leads to
some consequences which are nonintuitive to many who are first exposed
to its uses. In particular, since it is clearly possible for two
disjoint regions $V_1,V_2$ to be contained in the casual completion
$V''$ of a third region $V$, itself disjoint from $V_1 \cup V_2$,
it is possible for a single element $A \in \cA(V'')$ to be an element
of both $\cA(V_1 \cup V_2)$ and $\cA(V)$ and therefore to be localized
in mutually disjoint regions.\footnote{Indeed, this fact is essential
in our proof of Prop. \ref{main}.} Since the operational interpretation
of a self-adjoint $A \in \cA(V)$ is that of an observable measurable in
$V$, this leads to some initial conceptual discomfort.

     This discomfort is dissolved by noting that an ``observable'' $A$
does not represent a unique measuring apparatus in some fixed laboratory, but
rather represents an equivalence class of such apparata
(cf. \cite{Neumann-Werner}).  Consider two such idealized apparata
$X,Y$ such that $\phi(X) = \phi(Y)$ for all (idealized) states
$\phi$ admitted in the theory (the set of such states contains as
a subset --- at least in principle --- all states preparable in the
laboratory). These two apparata are then identified to be in the same
equivalence class and are thus represented by a single operator $A$.
Hence, the element $A$ above, which is localized
simultaneously in $V$ and $V_1 \cup V_2$, represents two distinct
events --- one taking place in $V$ and the other taking place in
$V_1 \cup V_2$. The fact that it is possible, given any event in
$V_1 \cup V_2$, to find an event in $V$ which is equivalent to the first
in the stated sense is part of the content of the local primitive causality
condition. It is therefore of interest that one can actually verify
this condition in models.

     Of further relevance to our purposes is the observation that the
use of local primitive causality leads to the conclusion that two
correlated projections $A,B$ yield an infinity of events, each of which is
localized in a manner disjoint from the others and is a common cause
of $A$ and $B$. Relativistic quantum field theory is extremely rich in
common causes!

     Proposition \ref{main} locates the common cause $C$ within
the union of the backward light cones of $V_1$ and $V_2$; however, a bit
more can be said of its location. Define $\tilde{V}_1$ and $\tilde{V}_2$ by
\begin{eqnarray}
\tilde{V}_1&\equiv& (BLC(V_1)\cap V)\setminus (BLC(V_1)\cap BLC(V_2)) \\
\tilde{V}_2&\equiv& (BLC(V_2)\cap V)\setminus (BLC(V_1)\cap BLC(V_2))
\end{eqnarray}
Since $(\tilde{V}_1\cup V_1)$ and $(\tilde{V}_2\cup V_2)$ are contained in
spacelike separated double cones, the algebras $\cN(\tilde{V}_1\cup V_1)$ and
$\cN(\tilde{V}_2\cup V_2)$ are logically independent, hence the common cause
$C< A\es B$ cannot belong to $\cN(\tilde{V}_1)$ or to $\cN(\tilde{V}_2)$
only, so neither $V\subseteq \tilde{V}_1$ nor $V\subseteq \tilde{V}_2$
is possible.

      Finally, we note that the existence of a (weak) common cause in
the presence of a violation of Bell's inequalities may seem
paradoxical, because the violation of Bell's inequalities is represented
by some (see, {\it e.g}, \cite{Fraassen}) as
implying the nonexistence of a common cause. But there is no
contradiction here --- it is essential to realize (cf. \cite{Redei1997}\cite{bjps})
that Bell's
inequality involves four pairs of correlated projections. To show that
Bell's inequality must hold, \cite{Fraassen} effectively assumes
that all pairs have the same common cause, {\it i.e.} a {\it common}
common cause $C$. We have demonstrated that a given pair of correlated
projections has a (weak) common cause, not that some set of four
correlated pairs has a common common cause. Common common causes for
different correlations do not exist in general even in classical
probability theory, as shown in \cite{HRS}.

\bigskip
\noindent {\bf Acknowledgement} This work was supported in part by
OTKA (contract numbers T 032771, T 035234, T 043642, T 037575 and
TS 04089).

\end{document}